\documentclass[fleqn,usenatbib]{mnras}

\usepackage{newtxtext,newtxmath}

\usepackage[T1]{fontenc}

\DeclareRobustCommand{\VAN}[3]{#2}
\let\VANthebibliography\thebibliography
\def\thebibliography{\DeclareRobustCommand{\VAN}[3]{##3}\VANthebibliography}

\usepackage{graphicx}	
\usepackage{amsmath}	
\usepackage{tabularx}

\usepackage{hyperref}

\newcommand{\logm}{\log_{10}(M_{\textrm{star}})}

\title[Quiescent galaxies across cosmic time]{Using the Star Forming Main Sequence To Explore Quiescent Galaxies Across Cosmic Time}

\author[T. Houston \& D. J. Croton \& M. Sinha]{
Tyler Houston,$^{1,2}$
Darren J. Croton$^{1,2}$\thanks{E-mail: dcroton@swin.edu.au}
Manodeep Sinha$^{1,2}$\\
$^{1}$Centre for Astrophysics \& Supercomputing, Swinburne University of Technology, Hawthorn, VIC 3122, Australia\\
$^{2}$ARC Centre of Excellence for All Sky Astrophysics in 3 Dimensions (ASTRO 3D)
}

\date{Accepted XXX. Received YYY; in original form ZZZ}
\pubyear{2022}

\begin{document}
\label{firstpage}
\pagerange{\pageref{firstpage}--\pageref{lastpage}}
\maketitle

\begin{abstract}
In this letter, we explore the quiescent lives of central galaxies using the SAGE galaxy model and Uchuu dark matter simulation. We ask three questions: (1) How much of a galaxy’s life is spent in quiescence? (2) How often do galaxies transition off the main sequence? (3) What is the typical duration of a quiescent phase? We find low and high-mass galaxies spend the highest fraction of their lives in quiescence: $45\pm19$\% for $\logm<9.0$ ($3.68\pm1.80$ Gyr) and $26\pm25$\% for $\logm>11.5$ ($3.46\pm3.30$ Gyr), falling to $7\pm13$\% for galaxies in-between ($0.82\pm1.57$ Gyr). Low mass galaxies move in and out of quiescence frequently, $2.8\pm1.3$ times on average, though only for short periods, $1.49\pm1.04$ Gyr. This can be traced to the influence of supernova feedback on their quite stochastic evolution. Galaxies of higher mass have fewer quiescent periods, $\sim0.7\pm0.9$, and their length increases with mass, peaking at $1.97\pm2.27$ Gyr. However, our high-mass population comprises star-forming and quiescent galaxies with diverging evolutionary paths, so the actual time may be even longer. These high-mass trends are driven by radio mode feedback from supermassive black holes, which, once active, tend to remain active for extended periods.
\end{abstract}




\section{Introduction}
\label{sec:intro}

The star forming main sequence relates the stellar mass of a galaxy to its star formation rate and is a relationship that describes the large-scale growth of the galaxy population. Similar correlations are seen in almost all areas of physics -- i.e. how a physical property relates to its time (or other) derivative. Analysing the evolution of the star forming main sequence over cosmic time provides a pathway to understanding how galaxy populations are born, evolve, and die.

Detailed measurements of the star forming main sequence only became possible once we entered the era of large-scale galaxy surveys. \cite{Noeske2007} found that at $z < 1.1$, $\log_{10}(M_{\textrm{star}})$ is proportional to $\log_{10}(\dot{M}_{\textrm{star}})$ with a limited range of star formation rates for a given $M_{\textrm{star}}$ \citep[see also][]{Elbaz2007}. Their work has since been expanded \citep[e.g.][]{Tomczak2016, Elbaz2018}, and the main sequence is now known to be present as early as $z \sim 6$ \citep{Speagle2014}. 

The relation clearly encodes many critical physical processes that shape galaxy evolution. The correlation itself appears to be set by the balance of gas supply into the galaxy and the efficiency with which this gas is converted into stars \citep{Kennicutt1998}. Galaxies above the main sequence are typically known to be undergoing interactions or mergers that perturb the gas and drive up the star formation rate. Galaxies below are usually experiencing short- or long-term episodes of star formation suppression \citep{Croton2006}. The latter can be caused by the removal of disk gas via feedback from supernova or active galactic nuclei and by limiting outside gas cooling into the galaxy. These processes determine how the main sequence changes with time, both in its normalisation and slope \citep{Speagle2014}. Thus, by measuring the main sequence, we gain insight into the background conditions for star formation across the galaxy population and the physical processes that govern them.

There is little consensus in the literature on the precise shape of the star forming main sequence. It has been measured to turn over at stellar masses $\logm > 10.5$ \citep{Lee2015, Leslie2020}. \cite{Thorne2021} show that this high mass turnover may only be present at low redshift ($z < 0.45$), finding no evidence of a turnover above $z = 2.6$. They postulate this is expected given that massive galaxies begin quenching more systematically at low redshift \citep{Katsianis2019}, which could be due to a growing bulge component that contributes to the stellar mass but not star formation rate. The scatter of the main sequence is of order $\sim 0.2 - 0.35$ dex \citep{Speagle2014, Tomczak2016}. \cite{Pearson2018} describe the small size of this scatter as being the result of every galaxy having star formation governed by similar quasi-static processes \citep{Lee2015}, with the scatter arising due to fluctuations in the mass flows into and within galaxies \citep{Mitra2017}. The normalisation of the main sequence has been broadly shown to increase with increasing redshift \citep{Tomczak2016, Pearson2018, Thorne2021}. 

Individual galaxies can be used as `test particles' to explore how a galaxy (star forming or quiescent) may evolve relative to the broader sequence. This evolution can tell us about the processes and events important for that galaxy’s history, even if we only have star formation and stellar mass information and nothing else. Unfortunately, this test can't be done observationally (at least not for the complete history of a galaxy), so one must resort to numerical models and simulations of galaxy evolution to map out the possibilities. Such models will be central to the analysis undertaken in this work.

In this letter, we ask three important questions about galaxy evolution relative to the star forming main sequence: (1) How much of a galaxy’s life is spent in quiescence? -- as defined by being found below the main sequence at any point in time. (2) How often do galaxies jump off and back on the main sequence? -- highlighting the stochastic nature of their growth. (3) What is the average time a galaxy is quiescent when off the main sequence? -- revealing the impact of the key physical processes on galaxy star formation. 

This letter is structured as follows: In Section~\ref{sec:model}, we describe the simulation and galaxy model used to build our data set. In Section~\ref{sec:method}, we outline our method to measure the star forming main sequence at each redshift and how galaxies move relative to it as they evolve across time. Section~\ref{sec:results} presents our results and the answers to the three questions given above, along with a brief discussion. We summarise in Section~\ref{sec:summary}. Throughout this work, we assume a Planck 2020 cosmology \citep{Aghanim2018} and a Hubble constant where $h = 0.73$. All stellar masses are given in units of solar mass, $M_\odot$.

\section{The Uchuu cosmological simulation and SAGE galaxy formation model}
\label{sec:model}

\begin{table*}
	\centering
	\begin{tabularx}{\linewidth}{X*{8}{c}}	
		$log_{10}(M_{star})$					&All			    &$8.5-9$	        &$9-9.5$		    &$9.5-10$            &$10-10.5$          &$10.5-11$         &$11-11.5$          &$11.5-12$\\
		\hline		
		\noalign{\vspace{4pt}}					
		Age	(Gyr)								&11.5 $\pm$ 0.8 	&8.09 $\pm$ 2.05    &10.5 $\pm$ 1.2	    &11.4 $\pm$ 0.9   	&12.0 $\pm$ 0.7     &12.5 $\pm$ 0.5    &12.9 $\pm$ 0.3     &13.2 $\pm$ 0.2   \\ \noalign{\vspace{4pt}}
		Time in Quiescence	(Gyr)				&2.18 $\pm$ 2.16	&3.68 $\pm$ 1.80    &1.84 $\pm$ 1.70	&0.82 $\pm$ 1.57	&1.27 $\pm$ 2.51    &1.87 $\pm$ 3.14   &2.38 $\pm$ 3.13    &3.46 $\pm$ 3.30   \\ \noalign{\vspace{4pt}}
        Quiescent Lifetime Fraction (\%)        &20 $\pm$ 20      &45 $\pm$ 19        &17 $\pm$ 16        &7 $\pm$ 13         &10 $\pm$ 20        &15 $\pm$ 24       &18 $\pm$ 24        &26 $\pm$ 25    \\ \noalign{\vspace{4pt}}
		Number of Quiescent Periods				&1.4 $\pm$ 1.1	&2.8 $\pm$ 1.3    &2.2 $\pm$ 1.4	&1.0 $\pm$ 1.0	&0.7 $\pm$ 0.9    &0.6 $\pm$ 0.8   &0.7 $\pm$ 0.9    &1.6 $\pm$ 1.5   \\ \noalign{\vspace{4pt}}
		Quiescent Period Length (Gyr)		    &1.32 $\pm$ 1.55	&1.49 $\pm$ 1.04    &0.80 $\pm$ 0.84	&0.53 $\pm$ 1.07	&0.99 $\pm$ 2.15    &1.54 $\pm$ 2.75   &1.97 $\pm$ 2.72   &1.98 $\pm$ 2.27   \\ \noalign{\vspace{4pt}}
        \noalign{\vspace{4pt}}
	\end{tabularx}
	\caption{\label{stats-table} The means and standard deviation of the quiescent galaxy population in our model, binned by $z = 0$ stellar mass. The columns show the age of galaxies, the time spent in quiescence, their lifetime fraction, the average number of quiescent periods, and the typical length of each. See also Fig. \ref{fig:LifetimeStatistics}.}
\end{table*}

Our work follows the semi-analytic approach to building a galaxy population. This is a hybrid method that evolves galaxies using an analytic model of the baryonic physics, post-processed on the output of a numerical simulation of the large-scale dark matter. 

For this dark matter backbone, we use the Uchuu Main simulation from the Uchuu Simulation Suite \citep{Ishiyama2021}. This suite comprises a series of $\Lambda$-cold dark matter N-body simulations covering various cosmological scales and mass resolutions. Uchuu Main, the flagship of the suite, was run using $12800^3$ dark matter particles in a box of side-length $2 h^{-1}$Gpc, and with a softening length of $4.27 h^{-1}$kpc. This resulted in a particle mass resolution of $3.27 \times 10^8 h^{-1} M_\odot$, leading to a minimum 20-particle resolved halo mass of $6.54 \times 10^9 h^{-1} M_\odot$. The simulation particles were evolved from $z = 127$ to $z = 0$ and produced 50 complete particle and halo data outputs to capture the evolution of structure at distinct times. These outputs serve as the input into our galaxy model. Uchuu was run using the latest (at the time) cosmological parameters measured by the Planck Satellite \citep{Aghanim2018}. For more information on Uchuu we refer the interested reader to \cite{Ishiyama2021}.

To populate Uchuu with galaxies we use the SAGE (Semi-Analytical Galaxy Evolution) galaxy formation model \citep{Croton2016}. SAGE describes the behaviour and evolution of baryonic matter in dark matter halos across cosmic time. This includes the hot, cold, and accreted gas captured by the halo, stars, and black holes that are formed from this gas by various processes, and ejected gas expelled from the halo due to strong feedback mechanisms. How the baryons in their various forms move between these mass reservoirs is governed by analytic prescriptions that model various relevant physical phenomena; these include cooling flows, star formation and metal production, and supernova and active galactic nuclei (AGN) feedback. The efficiency of such processes are governed by various parameters which give the modeller some control over how the galaxies evolve. In this letter, star formation and feedback are particularly important, as they are the prime influencers of the shape of the star forming main sequence. As such, below we provide a summary of their implementation in SAGE. Note that in this letter we use the SAGE 2016 model with default parameters. Full details can be found in \cite{Croton2016}.

Star formation in SAGE occurs when the mass of cold gas in a galaxy's disk exceeds a critical threshold. This indicates that the local gas density is sufficient for new stars to form, which is modelled using a Kennicutt-Schmidt-type star formation law \citep{Kennicutt1998}. Star formation continues while the cold gas mass remains above this threshold and uses up the gas, although it can be replenished from gas cooling out of the surrounding halo. The star formation rate of a galaxy in SAGE at any given time is simply the mass of cold gas in the disk above the critical threshold divided by the disk dynamical time, modulated by a conversion efficiency parameter that hides much of the (known and unknown, and unavoidable) smaller-scale temporal and spatial detail.

The cold disk gas, and hence star formation rate, is further modulated by various feedback mechanisms. SAGE models two kinds of feedback, one from supernova which heats and removes cold gas from the disk via supernova winds, and the other from AGN which heats and suppress the cooling of hot gas into the disk via quasar winds and radio jets. Both feedback mechanisms act to reduce the mass of cold gas available for star formation, and hence can influence a galaxy's place on the star forming main sequence. 

For supernova, the feedback rate is (of course) dependent on the presence of star formation. Thus, supernova winds tend to be most impactful in low-to-intermediate mass systems where star formation is common and the potential well of the parent dark matter halo is not so deep to escape. Collectively, supernova explosions across a galaxy heat and blow cold gas out into the halo surrounding the disk, and can even eject the gas out of the halo itself if the rate is high and halo potential shallow. The affected gas may later cool (for hot gas) or be reincorporated (for ejected gas) to once again form stars at a future time.

For AGN, SAGE differentiates two modes of feedback: the quasar mode and radio mode. The quasar mode refers to the strong hydrodynamic and radiative winds from optical/UV and X-Ray AGN \citep{Arav2001, Reeves2003}, triggered from galaxy--galaxy mergers and internal disk instabilities in the presence of cold disk gas. The disturbance drives this gas onto the central supermassive black hole, fuelling the wind from the surrounding accretion disk. Such winds act similarly to supernova but with significantly more energy; quasars in SAGE can remove the entire gas disk and hot halo. Hence, quasar winds are `event driven', with conditions that make them most common at high redshift in low-to-intermediate mass systems. 

In contrast, the radio mode of AGN results from the slow accretion of hot gas in the vicinity of the supermassive black hole, producing a low energy and (averaged over cosmological time-scales) continuous heating source into the galaxy and halo. SAGE assumes this heating interacts with the cooling gas from the hot halo, offsetting its energy losses and robbing the disk of fresh star forming material. For this to happen a massive and stable hot gas atmosphere must be present, and the black hole must be large enough to efficiently capture this gas within its gravitational radius. These conditions are typically met at late times in high mass systems when much of structure formation has settled. In this way, the radio mode is quite distinct to the quasar mode in both how it occurs, and its impact on galaxy evolution.

\section{Measuring the star forming main sequence across cosmic time}
\label{sec:method}

\begin{figure}
	\includegraphics[width=\columnwidth]{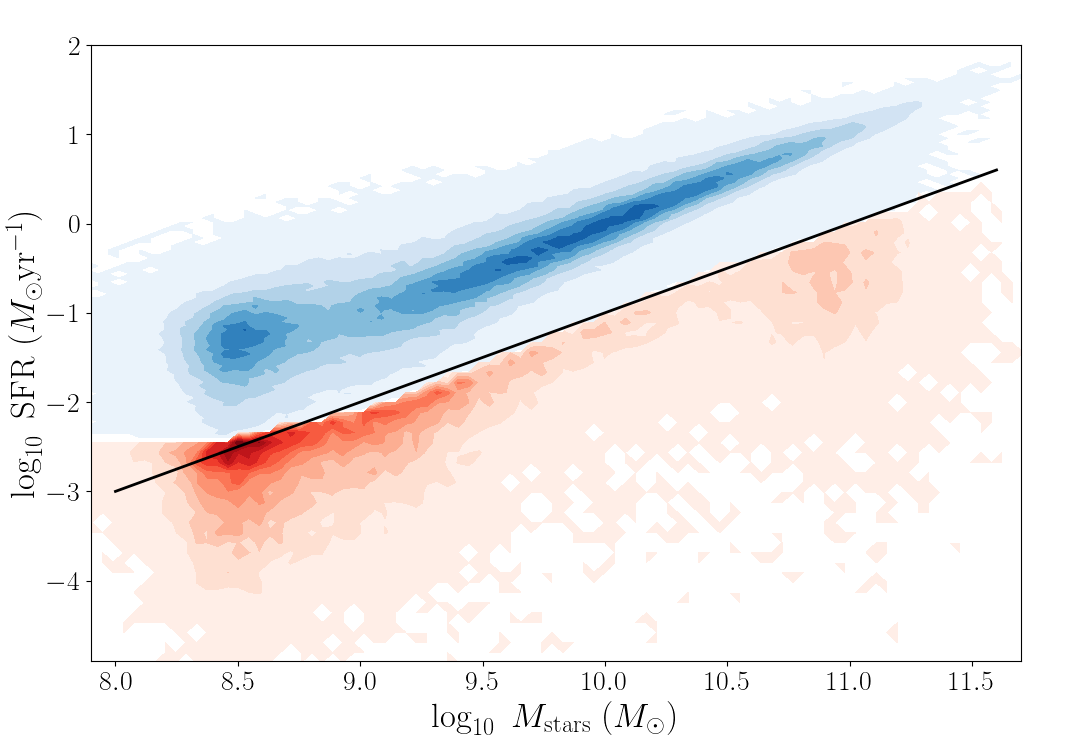}
    \caption{The star forming main sequence at $z = 0$ from the Uchuu N-body simulation and SAGE galaxy model. The contours show the star forming population in blue and quiescent population in red, with different normalisations for each to highlight their individual distributions. The solid black line marks the commonly used local observational selection $\log_{10}(\textrm{sSFR}) < -11$ to identify quiescent galaxies.}
    \label{fig:ContourSFQ}
\end{figure}

Once the galaxy formation model has run on the Uchuu simulation output, we have on-hand a vast data set that we can explore, approximately 4.4 million galaxies. Plotting each galaxy's star formation rate against its stellar mass at a given redshift is straightforward, but defining the star forming main sequence for the population as a whole, in a consistent way across all redshifts that can be fairly compared (out to those where galaxies are first born), is challenging. 

In this letter we adopt the method of \cite{Donnari2019}, whereby at a fixed redshift, the galaxy population is split into stellar mass bins with width 0.1 dex. The median star formation rate for each bin is then calculated iteratively, where for each iteration, galaxies with a star formation rate more than 1 dex below the median are defined as quiescent and removed. This continues until the median star formation rate for the remaining galaxies converges to 1\%. From this, galaxies with a star formation rate more than 1 dex below this converged median comprise the quiescent population, with the remainder making up the star forming population. This process is repeated for all 50 outputs of the SAGE+Uchuu data set (i.e. from $z=127$ to $z=0$) to find the star forming main sequence and quiescent galaxies in each.

The $z = 0$ outcome of this algorithm is presented in Figure~\ref{fig:ContourSFQ}. Here, the star forming main sequence is shown with blue contours, while galaxies defined as quiescent are shown with red contours. Note that because the quiescent population has much lower space density overall, we highlight their distribution by using a different contour normalisation to the star forming main sequence. 

This definition for quiescence also aligns with a common definition used in the observational literature for local data, where galaxies with a $\log_{10}$ specific star formation rate (sSFR) per year less than $-11$ are called quiescent \citep[e.g.][]{Ilbert2013}. This discriminator is shown in Figure~\ref{fig:ContourSFQ} with the solid black line. The advantage of the Donnari et al. method is that it can be applied out to high redshift and self-adjusts for changes in normalisation and slope, necessary for the goals of our work. 

Once done, every central $z = 0$ galaxy in our catalogue more massive than $\logm > 8.5$ is traced back in time along its main progenitor line to its snapshot of first identification in the simulation. This minimum value is chosen to stay well above the stellar mass completeness limit of the model, which is 0.5-1.0 dex lower. Our focus on central galaxies is due to their typically less complicated formation histories in comparison to satellite galaxies, and their more direct measurement and interpretation in the simulation. From the identification snapshot we linearly interpolate to find the galaxy's formation time, defined as the time since the Big Bang that this galaxy first reached $20$\% of its $z = 0$ mass. This chosen definition is both robust and reproducible (i.e. simulation resolution independent) \citep{Gao2005}. With this, and the galaxy's position relative to the star forming main sequence at each output, we can calculate the total time and fraction of its life the galaxy is quiescent, how many times it becomes quiescent since `formation', and the average time each quiescent period lasted. Specifically, a quiescent period is defined as a single snapshot or several consecutive snapshots where the galaxy is quiescent, and the quiescent time is simply the sum of those snapshot times. As such, our analysis is limited by the Uchuu snapshot time resolution of 200-400 Myr (depending on redshift), and we are smoothing over quiescent activity that occurs on shorter time-scales. Regardless, the results below give us confidence that this technical limitation is not biasing our conclusions. Finally, the number of quiescent periods a galaxy experiences is simply the number of times it has fallen off the main sequence since its formation time. Exploring the quiescent population statistics as a function of $z = 0$ stellar mass addresses the three questions we set as our goal at the outset of this letter. These we now discuss.

\section{Results: quantifying quiescence relative to the main sequence}
\label{sec:results}

\begin{figure}
	\includegraphics[width=\columnwidth]{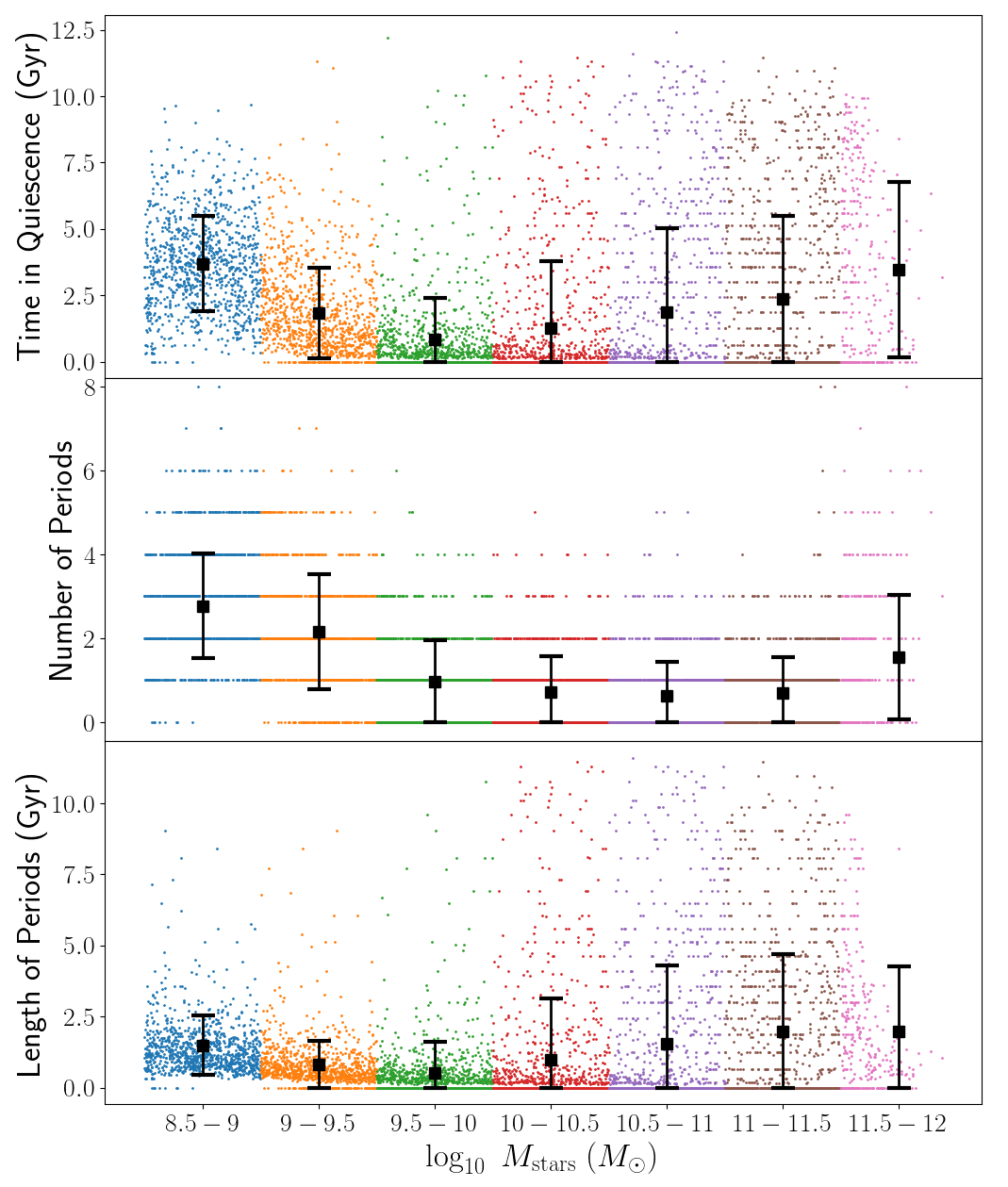}
    \caption{The properties of quiescent galaxies in the Uchuu N-body simulation and SAGE galaxy model shown as a function of $z = 0$ stellar mass. From top to bottom: the total time a galaxy is quiescent across its lifetime, the number of quiescent periods a galaxy experiences, and the average length of a quiescent period. Means and the $1\sigma$ variance for each mass range is shown by the symbols and bars. Up to 1000 randomly sampled individual galaxy data points from each mass bin are also over-plotted.}
    \label{fig:LifetimeStatistics}
\end{figure}

Table~\ref{stats-table} outlines our key results, which we also plot in Figure~\ref{fig:LifetimeStatistics} as a function of $z = 0$ stellar mass. In this figure, the three panels display (top) the number of years galaxies of each mass spend in quiescence across their life; (middle) the number of quiescent periods experienced by these galaxies; and (bottom) the length of these quiescent periods. For each mass bin, the average value is shown with error bars of one standard deviation. Up to 1000 data points were also sampled randomly from each range and are over-plotted to additionally highlight the change in the shape of the distribution with increasing stellar mass.

The top panel in Figure~\ref{fig:LifetimeStatistics} reveals that low mass ($\logm < 9.5$) and high mass ($\logm > 10.5$) galaxies spend more of their life in quiescence (up to $3.68$ Gyr) than galaxies of intermediate mass ($\logm \sim 10.0$, $0.82$ Gyr), on average. The $1\sigma$ range of ages increases significantly towards high mass, reflecting the more complicated and varied formation histories massive galaxies tend to experience. 

In the model we find this U-shaped trend with stellar mass reflects the influence of two primary physical processes: supernovae feedback and the radio mode of AGN. Lower mass galaxies have increasingly shallow gravitational potential wells which make their cold gas reservoirs susceptible to emptying via supernovae feedback. More massive galaxies often host AGN undergoing radio mode feedback due to the immense accretion occurring around their supermassive black hole. Both these feedback processes reduce the gas available for star formation and thus promote quiescence. Galaxies of intermediate mass fall between these extremes where such feedback effects are diminished. Hence, they tend to live more of their life happily forming stars on the star formation main sequence.

The middle and bottom panels of Figure~\ref{fig:LifetimeStatistics} support this view by showing the average number of quiescent periods and their average length when in quiescence, respectively. We see that low mass galaxies typically undergo more quiescent periods than higher mass galaxies, up to an average of $2.8$ times across their life. However these quiescent periods tend to be shorter, lasting on average $1.49$ Gyr. Again, this behaviour can be understood by the action of supernovae on the galaxy and its surrounds. Although the potential well of a galaxy can quickly be emptied of its gas from such feedback, it can similarly be refilled once the supernovae has passed, leading the galaxy back to star forming and the main sequence.

At intermediate-to-high galaxy masses the average number of quiescent periods is lower and levels out at $\sim 0.7$ times (noting a small up-tick for the most massive galaxies), and become longer. More massive systems are typically more stable owing to their deeper potential wells, and hence more significant disruptions are required to knock them off the star forming main sequence. The radio mode provides such disruption, suppressing the accretion of fresh gas for star formation. As such AGN typically represent a new phase in a galaxy's evolution and not something transient, radio AGN tend to affect the long-term evolution of a galaxy, unlike supernovae. Thus quiescence is less frequent (middle panel) but more impactful (bottom panel).

It is worth noting that one might expect the average length of a quiescent period for the most massive galaxies to be even longer than our results suggest, given that massive galaxies are known to be quenched for a significant fraction of their life. But our results measure the average population behaviour, which, in this mass range, includes a mix of large star forming spiral galaxies and massive quenched ellipticals (with an exponentially declining abundance). This mix of populations dilutes the effect somewhat. Here, it may be that the SAGE model is failing to capture the correct abundance massive quiescent galaxies at some level. Further exploration on this is required.

\begin{figure}
	\includegraphics[width=\columnwidth]{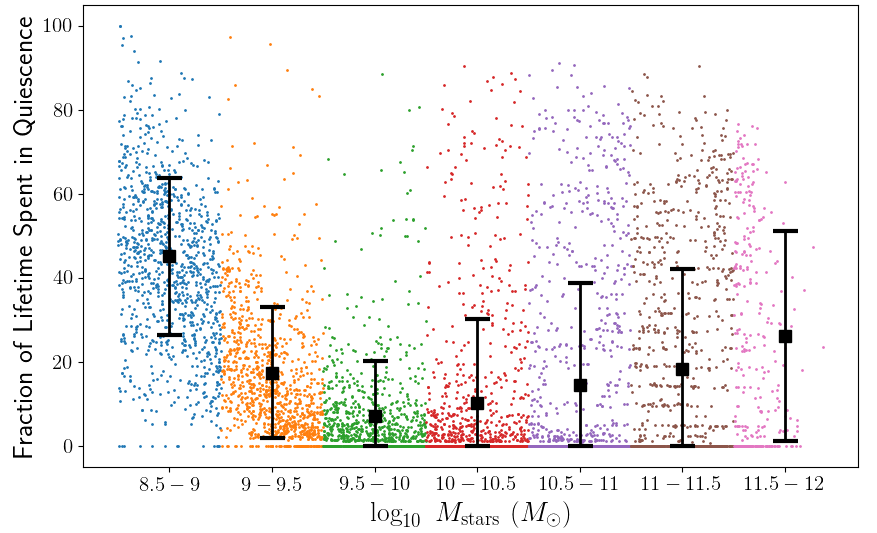}
    \caption{Similar to Figure~\ref{fig:LifetimeStatistics}, but showing the average fraction of a galaxy's total lifetime spent in quiescence, binned by stellar mass.}
    \label{fig:LifetimeFraction}
\end{figure}

Finally, we can also ask not just about the \textit{total} time a galaxy spends in quiescence, but the \textit{fraction} of its life. This is plotted in Figure~\ref{fig:LifetimeFraction}, following the same conventions as Figure~\ref{fig:LifetimeStatistics} (see also Table~\ref{stats-table}). We see that low and high mass galaxies live the largest fraction of their lives under the star forming main sequence, $45$\% and $26$\% respectively, but this happens for different reasons. Low mass galaxies at $z = 0$ are significantly younger on average (Table~\ref{stats-table}) and continually experience supernovae whenever they get the chance to grow via star formation (that then leads to short quiescent periods). High mass galaxies are almost always older and see a change in their population mix, with increasing mass having more radio mode quenched ellipticals, as discussed above. Galaxies in-between are the most efficient star formers in the Universe \citep{Behroozi2013}, and not surprisingly spend the lowest fraction of their lives under the main sequence, on average $7$\%.

\section{Summary}
\label{sec:summary}

In this letter we model the star forming main sequence using the Uchuu N-body cosmological dark matter simulation and the SAGE galaxy formation model. Central galaxies are tracked from their formation at high redshift to $z = 0$ in order to record their position relative to the main sequence throughout their lifetime. From this we quantify the amount of time galaxies spend on the main sequence as well as the number and duration of their quiescent periods, all as a function of final stellar mass. Our results are summarised in Table~\ref{stats-table} and Figure~\ref{fig:LifetimeStatistics} and \ref{fig:LifetimeFraction}.

Our results show that the low mass galaxy population, those with $\logm < 9.5$, experience frequent, short-term quiescent periods, indicative supernova feedback and its transient nature and impact. High mass galaxies, those with $\logm > 10.5$, spend a similar total lifetime in quiescence, although with fewer individual quiescent periods. This reflects the role of radio AGN in shutting down star formation, and highlights the long-term impact AGN have on galaxy evolution. Galaxies of intermediate mass are found to be less susceptible to supernovae and less likely to host radio AGN, and thus spend the least amount of time in quiescence of all the galaxies studied. 

Although this `story' of galaxy evolution is not new \citep[see][]{Baugh2006}, our work provides new utility by quantifying these very basic properties of galaxy quiescence across cosmic time in an established galaxy formation model. Furthermore, these methods can be applied to essentially any galaxy model or simulation for additional predictions and to map out the broader theoretical uncertainties. Regardless, using the star forming main sequence as a reference point to study quiescence with galaxy models offers a very direct way to compare and interpret the observations, which are getting increasingly detailed and sophisticated with time.

\section*{Acknowledgements}
\addcontentsline{toc}{section}{Acknowledgements}
We thank the referee, Neven Caplar, for their helpful comments which improved the quality of this letter. The authors acknowledge the support of the Australian Research Council Centre of Excellence for All-Sky Astrophysics in 3 Dimensions (ASTRO 3D), through project number CE170100013. We further thank the Instituto de Astrofisica de Andalucia (IAA-CSIC), Centro de Supercomputacion de Galicia (CESGA), and the Spanish academic and research network (RedIRIS) in Spain for hosting Uchuu DR1 and DR2 in the Skies \& Universes site for cosmological simulations. The Uchuu simulations were carried out on Aterui II supercomputer at Center for Computational Astrophysics, CfCA, of National Astronomical Observatory of Japan, and the K computer at the RIKEN Advanced Institute for Computational Science. The Uchuu DR1 and DR2 effort has made use of the skun@IAA$\_$RedIRIS and skun6@IAA computer facilities managed by the IAA-CSIC in Spain (MICINN EU-Feder grant EQC2018-004366-P).

\section*{Data Availability}

The data used in this letter are available in the Skies \& Universes online repository for cosmological simulations (Uchuu N-body Simulation; \href{http://www.skiesanduniverses.org/Simulations/Uchuu}{http://www.skiesanduniverses.org/Simulations/Uchuu}) and the Theoretical Astrophysical Observatory (SAGE galaxy model; \href{https://tao.asvo.org.au}{https://tao.asvo.org.au}). SAGE is a publicly available code-base and can be download at \href{https://github.com/darrencroton/sage}{https://github.com/darrencroton/sage}. 



\bibliographystyle{mnras}
\bibliography{papers}

\begin{thebibliography}{}
\makeatletter
\relax
\def\mn@urlcharsother{\let\do\@makeother \do\$\do\&\do\#\do\^\do\_\do\%\do\~}
\def\mn@doi{\begingroup\mn@urlcharsother \@ifnextchar [ {\mn@doi@}
  {\mn@doi@[]}}
\def\mn@doi@[#1]#2{\def\@tempa{#1}\ifx\@tempa\@empty \href
  {http://dx.doi.org/#2} {doi:#2}\else \href {http://dx.doi.org/#2} {#1}\fi
  \endgroup}
\def\mn@eprint#1#2{\mn@eprint@#1:#2::\@nil}
\def\mn@eprint@arXiv#1{\href {http://arxiv.org/abs/#1} {{\tt arXiv:#1}}}
\def\mn@eprint@dblp#1{\href {http://dblp.uni-trier.de/rec/bibtex/#1.xml}
  {dblp:#1}}
\def\mn@eprint@#1:#2:#3:#4\@nil{\def\@tempa {#1}\def\@tempb {#2}\def\@tempc
  {#3}\ifx \@tempc \@empty \let \@tempc \@tempb \let \@tempb \@tempa \fi \ifx
  \@tempb \@empty \def\@tempb {arXiv}\fi \@ifundefined
  {mn@eprint@\@tempb}{\@tempb:\@tempc}{\expandafter \expandafter \csname
  mn@eprint@\@tempb\endcsname \expandafter{\@tempc}}}

\bibitem[\protect\citeauthoryear{{Arav}, {de Kool}  \& {Korista}}{{Arav}
  et~al.}{2001}]{Arav2001}
{Arav} N.,  {de Kool} M.,   {Korista} K.~T.,  2001, arXiv e-prints, \href
  {https://ui.adsabs.harvard.edu/abs/2001astro.ph..7415A} {pp
  astro--ph/0107415}

\bibitem[\protect\citeauthoryear{{Baugh}}{{Baugh}}{2006}]{Baugh2006}
{Baugh} C.~M.,  2006, \mn@doi [Reports on Progress in Physics]
  {10.1088/0034-4885/69/12/R02}, \href
  {https://ui.adsabs.harvard.edu/abs/2006RPPh...69.3101B} {69, 3101}

\bibitem[\protect\citeauthoryear{{Behroozi}, {Wechsler}  \&
  {Conroy}}{{Behroozi} et~al.}{2013}]{Behroozi2013}
{Behroozi} P.~S.,  {Wechsler} R.~H.,   {Conroy} C.,  2013, \mn@doi [\apj]
  {10.1088/0004-637X/770/1/57}, \href
  {https://ui.adsabs.harvard.edu/abs/2013ApJ...770...57B} {770, 57}

\bibitem[\protect\citeauthoryear{{Croton} et~al.,}{{Croton}
  et~al.}{2006}]{Croton2006}
{Croton} D.~J.,  et~al., 2006, \mn@doi [\mnras]
  {10.1111/j.1365-2966.2005.09675.x}, \href
  {https://ui.adsabs.harvard.edu/abs/2006MNRAS.365...11C} {365, 11}

\bibitem[\protect\citeauthoryear{{Croton} et~al.,}{{Croton}
  et~al.}{2016}]{Croton2016}
{Croton} D.~J.,  et~al., 2016, \mn@doi [\apjs] {10.3847/0067-0049/222/2/22},
  \href {https://ui.adsabs.harvard.edu/abs/2016ApJS..222...22C} {222, 22}

\bibitem[\protect\citeauthoryear{{Donnari} et~al.,}{{Donnari}
  et~al.}{2019}]{Donnari2019}
{Donnari} M.,  et~al., 2019, \mn@doi [\mnras] {10.1093/mnras/stz712}, \href
  {https://ui.adsabs.harvard.edu/abs/2019MNRAS.485.4817D} {485, 4817}

\bibitem[\protect\citeauthoryear{{Elbaz} et~al.,}{{Elbaz}
  et~al.}{2007}]{Elbaz2007}
{Elbaz} D.,  et~al., 2007, \mn@doi [\aap] {10.1051/0004-6361:20077525}, \href
  {https://ui.adsabs.harvard.edu/abs/2007A&A...468...33E} {468, 33}

\bibitem[\protect\citeauthoryear{{Elbaz} et~al.,}{{Elbaz}
  et~al.}{2018}]{Elbaz2018}
{Elbaz} D.,  et~al., 2018, \mn@doi [\aap] {10.1051/0004-6361/201732370}, \href
  {https://ui.adsabs.harvard.edu/abs/2018A&A...616A.110E} {616, A110}

\bibitem[\protect\citeauthoryear{{Gao}, {Springel}  \& {White}}{{Gao}
  et~al.}{2005}]{Gao2005}
{Gao} L.,  {Springel} V.,   {White} S. D.~M.,  2005, \mn@doi [\mnras]
  {10.1111/j.1745-3933.2005.00084.x}, \href
  {https://ui.adsabs.harvard.edu/abs/2005MNRAS.363L..66G} {363, L66}

\bibitem[\protect\citeauthoryear{{Ilbert} et~al.,}{{Ilbert}
  et~al.}{2013}]{Ilbert2013}
{Ilbert} O.,  et~al., 2013, \mn@doi [\aap] {10.1051/0004-6361/201321100}, \href
  {https://ui.adsabs.harvard.edu/abs/2013A&A...556A..55I} {556, A55}

\bibitem[\protect\citeauthoryear{{Ishiyama} et~al.,}{{Ishiyama}
  et~al.}{2021}]{Ishiyama2021}
{Ishiyama} T.,  et~al., 2021, \mn@doi [\mnras] {10.1093/mnras/stab1755}, \href
  {https://ui.adsabs.harvard.edu/abs/2021MNRAS.506.4210I} {506, 4210}

\bibitem[\protect\citeauthoryear{{Katsianis} et~al.,}{{Katsianis}
  et~al.}{2019}]{Katsianis2019}
{Katsianis} A.,  et~al., 2019, \mn@doi [\apj] {10.3847/1538-4357/ab1f8d}, \href
  {https://ui.adsabs.harvard.edu/abs/2019ApJ...879...11K} {879, 11}

\bibitem[\protect\citeauthoryear{{Kennicutt}}{{Kennicutt}}{1998}]{Kennicutt1998}
{Kennicutt} Robert~C. J.,  1998, \mn@doi [\apj] {10.1086/305588}, \href
  {https://ui.adsabs.harvard.edu/abs/1998ApJ...498..541K} {498, 541}

\bibitem[\protect\citeauthoryear{{Lee} et~al.,}{{Lee} et~al.}{2015}]{Lee2015}
{Lee} N.,  et~al., 2015, \mn@doi [\apj] {10.1088/0004-637X/801/2/80}, \href
  {https://ui.adsabs.harvard.edu/abs/2015ApJ...801...80L} {801, 80}

\bibitem[\protect\citeauthoryear{{Leslie} et~al.,}{{Leslie}
  et~al.}{2020}]{Leslie2020}
{Leslie} S.~K.,  et~al., 2020, \mn@doi [\apj] {10.3847/1538-4357/aba044}, \href
  {https://ui.adsabs.harvard.edu/abs/2020ApJ...899...58L} {899, 58}

\bibitem[\protect\citeauthoryear{{Mitra}, {Dav{\'e}}, {Simha}  \&
  {Finlator}}{{Mitra} et~al.}{2017}]{Mitra2017}
{Mitra} S.,  {Dav{\'e}} R.,  {Simha} V.,   {Finlator} K.,  2017, \mn@doi
  [\mnras] {10.1093/mnras/stw2527}, \href
  {https://ui.adsabs.harvard.edu/abs/2017MNRAS.464.2766M} {464, 2766}

\bibitem[\protect\citeauthoryear{{Noeske} et~al.,}{{Noeske}
  et~al.}{2007}]{Noeske2007}
{Noeske} K.~G.,  et~al., 2007, \mn@doi [\apjl] {10.1086/517926}, \href
  {https://ui.adsabs.harvard.edu/abs/2007ApJ...660L..43N} {660, L43}

\bibitem[\protect\citeauthoryear{{Pearson} et~al.,}{{Pearson}
  et~al.}{2018}]{Pearson2018}
{Pearson} W.~J.,  et~al., 2018, \mn@doi [\aap] {10.1051/0004-6361/201832821},
  \href {https://ui.adsabs.harvard.edu/abs/2018A&A...615A.146P} {615, A146}

\bibitem[\protect\citeauthoryear{{Planck Collaboration} et~al.,}{{Planck
  Collaboration} et~al.}{2020}]{Aghanim2018}
{Planck Collaboration} et~al., 2020, \mn@doi [\aap]
  {10.1051/0004-6361/201833910}, \href
  {https://ui.adsabs.harvard.edu/abs/2020A&A...641A...6P} {641, A6}

\bibitem[\protect\citeauthoryear{{Reeves}, {O'Brien}  \& {Ward}}{{Reeves}
  et~al.}{2003}]{Reeves2003}
{Reeves} J.~N.,  {O'Brien} P.~T.,   {Ward} M.~J.,  2003, \mn@doi [\apjl]
  {10.1086/378218}, \href
  {https://ui.adsabs.harvard.edu/abs/2003ApJ...593L..65R} {593, L65}

\bibitem[\protect\citeauthoryear{{Speagle}, {Steinhardt}, {Capak}  \&
  {Silverman}}{{Speagle} et~al.}{2014}]{Speagle2014}
{Speagle} J.~S.,  {Steinhardt} C.~L.,  {Capak} P.~L.,   {Silverman} J.~D.,
  2014, \mn@doi [\apjs] {10.1088/0067-0049/214/2/15}, \href
  {https://ui.adsabs.harvard.edu/abs/2014ApJS..214...15S} {214, 15}

\bibitem[\protect\citeauthoryear{{Thorne} et~al.,}{{Thorne}
  et~al.}{2021}]{Thorne2021}
{Thorne} J.~E.,  et~al., 2021, \mn@doi [\mnras] {10.1093/mnras/stab1294}, \href
  {https://ui.adsabs.harvard.edu/abs/2021MNRAS.505..540T} {505, 540}

\bibitem[\protect\citeauthoryear{{Tomczak} et~al.,}{{Tomczak}
  et~al.}{2016}]{Tomczak2016}
{Tomczak} A.~R.,  et~al., 2016, \mn@doi [\apj] {10.3847/0004-637X/817/2/118},
  \href {https://ui.adsabs.harvard.edu/abs/2016ApJ...817..118T} {817, 118}

\makeatother
\end{thebibliography}


\bsp	
\label{lastpage}
\end{document}